\documentclass[11pt,a4paper,onecolumn]{article} 
\usepackage{opex3}
\usepackage{amsbsy,amsmath,amssymb}
\usepackage{setspace}
\usepackage{upgreek}
\usepackage{color}
\usepackage{graphicx}

\begin{document}

\title{Position measurement of non-integer OAM beams with structurally invariant propagation}

\author{A. M. Nugrowati, W. G. Stam and J. P. Woerdman}
\address{\footnotesize{Huygens Laboratory, Leiden University, P.O. Box 9504, 2300 RA Leiden, NL\\
$^*$Corresponding author: nugrowati@physics.leidenuniv.nl}}

\begin{abstract} We present a design to generate structurally propagation invariant light beams carrying non-integer orbital angular momentum (OAM) using Hermite-Laguerre-Gaussian (HLG) modes. Different from previous techniques, the symmetry axes of our beams are fixed when varying the OAM; this  simplifies the calibration technique for beam positional measurement using a quadrant detector. We have also demonstrated analytically and experimentally that both the OAM value and the HLG mode orientation play an important role in the quadrant detector response. The assumption that a quadrant detector is most sensitive at the beam center does not always hold for anisotropic beam profiles, such as HLG beams.\end{abstract}


\bibliographystyle{osajnl}

\section{Introduction}
Light carrying orbital angular momentum (OAM) is characterized by a helical wavefront shape and a doughnut-like intensity profile with a dark center (vortex). In a single round trip about the propagation axis, the phase of an OAM beam increases linearly and gains the value of $2\pi N$, with $N$ an integer value that is equivalent to the OAM content of such a beam. After the first investigation of the astigmatic transformation of Hermite-Gaussian (HG) modes into Laguerre-Gaussian (LG) modes  \cite{Abramochkin:OptCommun1991}, it was theoretically proven that LG laser modes carry a well defined OAM which is equivalent to the azimuthal mode index $\ell$ of the LG modes \cite{Woerdman:PRA1992}. Since then, the generation of LG modes has opened up a broad range of applications, including optical trapping with OAM beam structures \cite{Dunlop:JMod1995, Padgett:NatPhoton2011}, quantum communication at higher dimensional entanglement using OAM beams \cite{Oemrawsingh:PRL2005, Pors:JOpt2011}, OAM beam for high sensitivity Raman spectroscopy in molecule detection \cite{Sato:OptCommun2007}, stellar detection using OAM beam \cite{Berkhout:OpEx2010, Tamburini:NatPhys2011}, and nanometer precision metrology by using the effect of OAM on beam shifts \cite{Hermosa:PRA2010}. 

Recently, there is a growing interest in addressing \emph{non-integer} values of OAM that potentially broadens the OAM beams applications. This will be the topic of our paper. We present the technique to generate non-integer OAM beams and discuss  the difference between our technique and the existing ones. Subsequently, we treat the position measurement of such a beam that is an inherent part of many applications using OAM beams.

During the first decade after the initial realization of an OAM beam, many different \emph{integer} OAM beam generation techniques have been introduced. The first demonstration used the so-called `$\pi/2$-mode converter', which belongs to a family of astigmatic mode converters that applies the appropriate Gouy phase to create well defined mode indices of LG beams carrying integer OAM \cite{Woerdman:OptCommun1993}. This was soon followed by the demonstration of a spiral phase plate (SPP) operating at optical wavelength \cite{Woerdman:OptCommun1994} and at milimeter range \cite {Padgett:OptCommun1996} for creating helical-wavefront to directly transform Gaussian beam to OAM beams. At the same time, computer-generated holograms with pitchfork structures were applied using a spatial light modulator (SLM) to convert Gaussian beams into LG beams \cite{Dunlop:JMod1995, Padgett:JMod1998}. Different from the astigmatic mode converter, both SPP and SLM are not pure mode converters. They convert a fundamental Gaussian mode into a superposition of LG modes that contain the same azimuthal mode index $\ell$ but different radial mode index $p$. Although the OAM content of such a beam is well defined, the spatial field distribution evolves during propagation. This mode impurity problem holds also when employing q plates \cite{Marucci:PRL2006} that convert spin-to-orbital angular momentum in an anisotropic and inhomogeneous media to create helical waves. Mitigating the radial mode impurity to obtain a more robust beam profile during propagation when using SLM and SPP has then been the focus of several studies \cite{Friesem:OptCommun2002, Abraham:PRA2002, Otsuka:OptCommun2008, Hara:JOSAA2008, Fukuchi:OptLett2009}.

In the field of \emph{non-integer} OAM beam generation, only a handful of studies have been carried on. One of the initial ideas was to use off-axis illumination of an SPP; equivalently, one may use a non-integer $2\pi$ phase step SPP \cite{Oemrawsingh:PRL2005}. These techniques, however, yield non-integer OAM beams with neither $\ell$ nor $p$ mode purity \cite{Berry:JOA2004}. A more structurally stable non-integer OAM beam has been demonstrated recently, by using an SLM when applying a synthesis of a finite number of LG modes with carefully chosen Gouy phases \cite{Gotte:OpEx2008}. It was, however, demonstrated only for half-integer OAM values. Another proposition is by exploiting the internal conical diffraction where a circularly polarized beam with a fundamental Gaussian mode is converted into a non-integer OAM beam with a Bessel mode, having only a limited OAM value range of $|\ell|\leq1$ \cite{Dwyer:OpEx2010}. 

Our paper focuses on two issues. The first concerns with the generation of beams carrying arbitrary non-integer OAM values that is structurally stable during propagation. This can be achieved by employing the concept of generally astigmatic mode converters, as was initially introduced in Ref.~\cite{Courtial:OptCommun2000}. Later on, it was theoretically demonstrated that the output of a general astigmatic transformation is the intermediate beam between HG and LG beams, known as Hermite-Laguerre-Gaussian (HLG) beams \cite{Abramochkin:JOA2004, Nienhuis:PRA2004, Abramochkin:JOSAA2010}. Since such a HLG mode is an analytic interpolation between a HG mode and a LG mode, it is structurally propagation invariant. Moreover, this HLG beam carries non-integer OAM.

The second issue concerns with the positional detection of generated HLG modes for applications of non-integer OAM beams, e.g. in precision metrology, optical tweezing or scanning near-field optical microscopy. Popular device to measure beam position is a quadrant detector, made of 2 by 2 array of photodiodes that are equally spaced and produce four electronic signals that are proportional to the beam position. The sensitivity of the position measurement using a quadrant detector is limited within a small spatial range, where the detector response to the shift of the beam is linear. This holds even for the position detection of a typical fundamental Gaussian beam. The non-linear response of a quadrant detector has been addressed and improved, but only for fundamental Gaussian beams \cite{Kouh:CurrApplPhys2010, Robinson:ApplOpt2010}. Recently, we have also investigated the response and correspondingly the calibration of a quadrant detector for LG modes carrying \emph{integer} OAM beams \cite{Hermosa:OptLett2011}. In this paper, we are going to discuss the use of a quadrant detector for beam positional detection \emph{non-integer} OAM beams based upon HLG modes.

We start our paper by presenting the design of a mode converter which transforms a HG mode of arbitrarily high order to a HLG mode \cite{Abramochkin:JOA2004}. We demonstrate that the output beam is structurally propagation invariant, and characterize the non-integer OAM value using an interferometer set-up. Different from the previous designs \cite{Courtial:OptCommun2000, Abramochkin:JOA2004}, our experimental set-up generates HLG beam with fixed symmetry axes when varying the non-integer OAM; these axes thus also overlap with the quadrant detector measurement axes. This proves to be very beneficial when detecting the beam position of HLG modes as shown in the second part of the paper. Further, we derive an analytical expression of a quadrant detector response towards general astigmatic modes and introduce a calibration procedure required for detecting the positional shifts of non-integer OAM beams.

\section{Generation of structurally propagation invariant light carrying non-integer OAM}
\subsection{Experimental set-up}
\begin{figure}[htbp]
\begin{center}
\includegraphics[width=11cm]{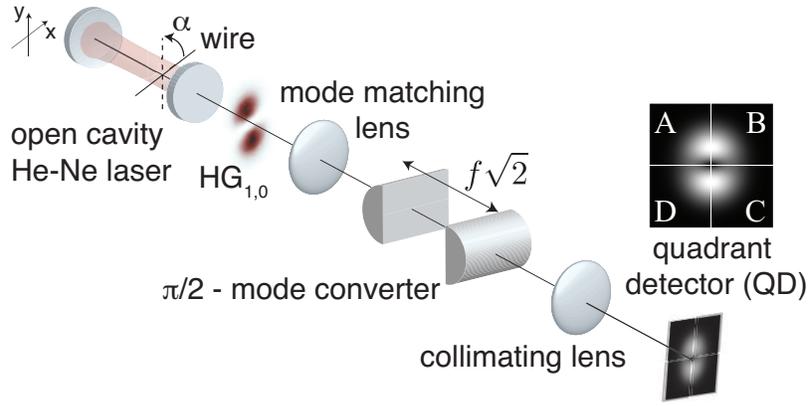}
\end{center}
\caption{\label{fig:set-up} Our experimental set-up to generate HLG modes as non-integer OAM beams, equipped with a quadrant detector for measuring the beam positional shifts, discussed in Section~\ref{Sec:QD}.}
\end{figure}
A conventional $\pi/2$ astigmatic mode converter \cite{Woerdman:OptCommun1993} transforms a pure HG mode into a pure LG mode by passing an incoming HG beam through a pair of identical cylindrical lenses with focal lengths $f$, separated at a distance $d=f\sqrt2$, as illustrated in Figure~\ref{fig:set-up}. A mode matching lens is normally used to tailor the beam waist of the outgoing laser mode into the desired beam waist in between the cylindrical lens. A well defined \textit{integer} OAM is achieved when the symmetry axes of the HG beam are oriented at an angle $\alpha=45^\mathrm{o}$ with respect to the symmetry axes of the cylindrical lenses \cite{Woerdman:OptCommun1993}. This can be done using an open laser cavity that is forced to operate at a high order HG mode by insertion of a thin metal wire, oriented at $\alpha=45^\mathrm{o}$. 
\begin{figure}[htbp]
\begin{center}
\includegraphics[width=11cm]{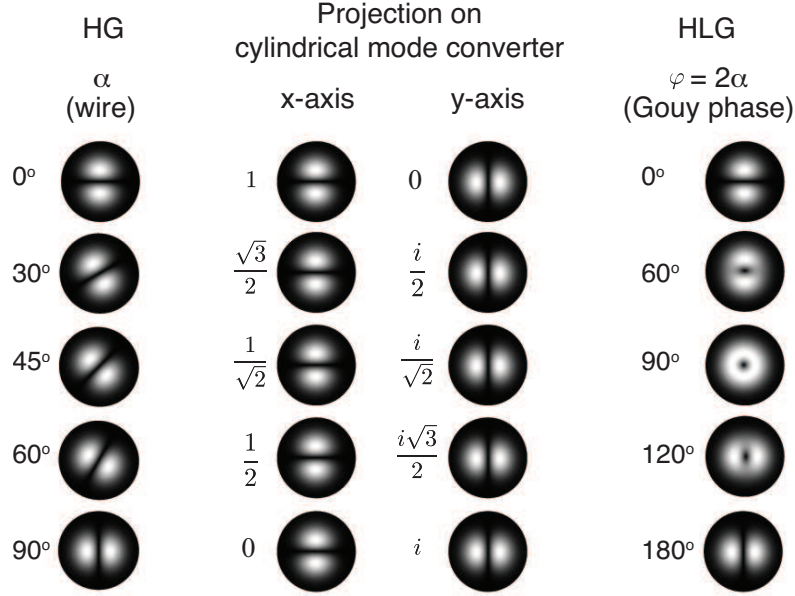}
\end{center}
\caption{\label{fig:proj} An incoming HG$_{0,1}$ mode at varying orientation angle $\alpha$ projected onto the orthogonal symmetry axes of the cylindrical lenses of a `$\pi/2$-mode converter'. The symmetry axes of the outgoing HLG modes are always aligned to the projection axes. The outgoing HLG modes experience Gouy phase $\varphi=2\alpha$.}
\end{figure}

Belonging to the family of astigmatically transformed HG beams, HLG beams can be created by tuning the beam parameter $\alpha$, i.e. the angle between the symmetry axes of cylindrical lenses and the symmetry axes of the input HG beam \cite{Abramochkin:JOSAA2010}. The non-integer OAM value of HLG beam is $\ell=(n-m) \sin 2\alpha$, with $n$ and $m$ the mode index of high order HG beams \cite{Abramochkin:JOA2004, Nienhuis:PRA2004}. Another way to generate HLG beams is to tailor the required Gouy phase by simultaneously tuning the separation distance of the cylindrical lenses $d$ and the position of the cylindrical lenses pair with respect to the mode matching lens \cite{Habraken:note}. However, for aligning purposes, the approach of varying $\alpha$ is more attractive when tuning the non-integer OAM value. 

A general astigmatic mode converter transforms a HG mode of arbitrarily high order to a HLG mode. In essence, a pure mode transformation projects an incoming HG mode into two orthogonal axes of the astigmatic mode converter. The outgoing HLG beam is a superposition of the projected mode with the additional Gouy phase. In Figure~\ref{fig:proj}, we show the projection of an incoming HG$_{01}$ mode with varying orientation angle $\alpha$ on a `$\pi/2$-mode converter'. The Gouy phase $\varphi$ experienced by the projected mode after traversing the  cylindrical lenses is $2\alpha$. Note that the symmetry axes of the outgoing HLG beam are always aligned to the projection axes of the cylindrical lenses. 

In this paper, we demonstrate the non-integer OAM beam generation using a HeNe gain tube (Spectra Physics 120S) operating at a wavelength $\lambda=632.8$~nm, situated at the centre of an open two-mirror cavity allowing for a generation of up to the third order of the HG mode family (i.e. HG$_{3,3}$). The laser is forced to operate in a single higher order HG mode by insertion of a $18~\mu$m diameter copper wire normal to and rotatable with respect to the axis of the laser cavity. The strength and location of a mode matching lens and a pair of cylindrical lenses are chosen such that they create integer OAM beams when the wire is orientated at $\alpha=45^\mathrm{o}$. By rotating the wire about the optical axis, we tune the parameter $\alpha$ to generate the HLG modes. This is different from the two previous techniques; where two Dove prisms and two cylindrical lenses are rotated to flip the HG mode before being converted into HLG modes \cite{Courtial:OptCommun2000}, or where $\alpha$ is tuned by rotating the cylindrical lenses \cite{Abramochkin:JOA2004}. By rotating the metal wire inside the open laser cavity in Figure~\ref{fig:set-up}, our technique generates HLG modes with a fixed symmetry axes for any arbitrary non-integer OAM value. Therefore, the profile mode axes are \textit{always} aligned to the quadrant detector measurement axes which greatly simplifies the quadrant detector operation.

\subsection{Characterization of non-integer OAM beams}
\begin{figure}[htbp]
\begin{center}
\includegraphics[width=13cm]{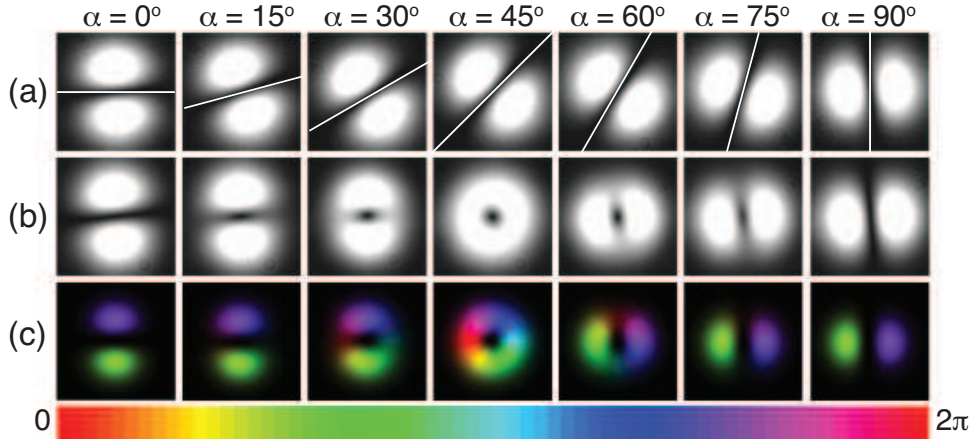}
\end{center}
\caption{\label{fig:HLGbeams} [Color online] (a) Measured intensity profiles of the impinging HG$_{0,1}$ mode as a function of the orientation angle $\alpha$ with respect to the symmetry axes of the `$\pi/2$-mode converter'; the white lines correspond to the wire orientation in the open laser cavity of Figure~\ref{fig:set-up}. (b) Measured far-field intensity profiles after the collimating lens of the outgoing Hermite-Laguerre-Gauss (HLG) modes. (c) Calculated intensity profiles to compare with the measurement results in (b). The color map in the calculated intensity profiles (c) corresponds to the HLG phase profile that gradually increases from 0 to $2\pi$.}
\end{figure}

\begin{figure}[htbp]
\begin{center}
\includegraphics[width=8.5cm]{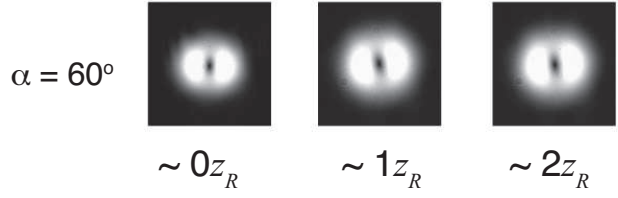}
\end{center}
\caption{\label{fig:HLGevolve} Mode profiles at several Rayleigh distance $z_\mathrm{R}$, representing the near- and far-field planes for the outgoing HLG$_{0,1|[\alpha:60^\mathrm{o}]}$ mode.}
\end{figure}

Figure~\ref{fig:HLGbeams} shows the resulting generated HLG beams as a function of varying orientation angle $\alpha$. The open laser cavity is forced to operate at the first higher order HG mode, i.e HG$_{0,1}$. The first two rows display the measured intensity profiles of (a) the incoming HG$_{0,1}$ and (b) the outgoing HLG$_{0,1|[\alpha:0^\mathrm{o},90^\mathrm{o}]}$ beams at the far-field after the collimating lens. Our generated HLG beam profiles match with the calculation shown in Figure~\ref{fig:HLGbeams}(c). In the calculated images, we have used a color map to indicate the phase profile of the generated HLG modes. For outgoing HLG profiles being the analytic interpolation between a HG mode and a LG mode, we observe a more flat wavefront inside the high intensity areas (note the even color tone). Inside the dark intensity areas, the phase value increases non-linearly along the azimuthal direction. The phase singularity of zero-OAM beams at $\alpha=N\times90^\mathrm{o}$ forms a line (most left and most right images of Figure~\ref{fig:HLGbeams}(c)), whereas for integer OAM beams at $\alpha=(2N+1)\times45^\mathrm{o}$ it forms a vortex (the center image of Figure~\ref{fig:HLGbeams}(c)), with $N$ an integer number. Apart from the overall scaling and slight astigmatism due to imperfect alignment, the generated HLG beams are structurally stable during propagation, as shown in Figure~\ref{fig:HLGevolve}. 


\begin{figure}[htbp]
\begin{center}
\includegraphics[width=12.5cm]{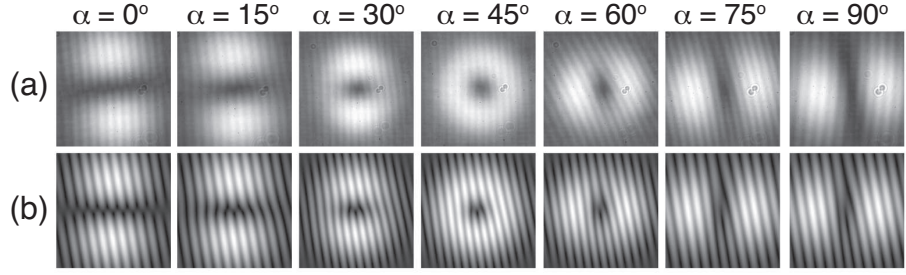}
\end{center}
\caption{\label{fig:HLGint} The (a) measured and (b) calculated interference patterns showing the phase singularity of the HLG beams. The black color corresponds to zero intensity and zero phase, whereas the white color corresponds to maximal intensity and phase $\phi=2\pi$.}
\end{figure}

To characterize the OAM content of the generated HLG beams, we look at the interference patterns between the outgoing HLG beam and a reference beam that comes out of the laser cavity. A typical interference pattern of an integer OAM beam shows phase dislocation features, i.e. a pitchfork that branches out into $\ell$ number of lines at the dark centre of the beam, which is also the case for the centre image of Figure~\ref{fig:HLGint}. In the case of beams with non-integer OAM values, the branching gradually dissolves into separated shifted lines as shown by the measurement result (a) and confirmed by the calculation (b). 

The perfect match between our measurement results and calculation, for both the intensity and the interference profiles, demonstrates that non-integer OAM values do indeed depend on the orientation angle of the incoming HG beam, expressed as $\ell = (n-m)\sin 2\alpha$ \cite{Abramochkin:JOA2004, Nienhuis:PRA2004}. There are two consequences of this relation when generating HLG beams using our set-up. First, the sign of $\ell$ changes each time $\alpha$ crosses the value of $N\times90^\mathrm{o}$. Secondly, the HLG mode profile rotates by $90^\mathrm{o}$ each time $\alpha$ crosses the value of $(2N+1)\times45^\mathrm{o}$ with $N$ an integer number.
Take the example of Figure~\ref{fig:HLGbeams}(b)-(c), where we have tuned the angle $0^\mathrm{o}\leq\alpha\leq90^\mathrm{o}$ to obtain $0\leq\ell\leq1$. Although all the HLG modes shown have positive values of $\ell$, the mode profiles for $\alpha\leq45^\mathrm{o}$ are rotated $90^\mathrm{o}$ with respect to the profiles for $\alpha\geq45^\mathrm{o}$. Therefore, for an identical $\ell$ or OAM values, there are two possible orthogonal orientations of the HLG beams. The orthogonal orientation of HLG beams also greatly simplifies the calibration procedure when measuring the beam position using a quadrant detector, as will be discussed in the next section. 

\section{Quadrant detector response to HLG beam displacement}\label{Sec:QD}
In this section we deal with the response of a quadrant detector as a beam positional  detector of HLG beams. A quadrant detector is a 2x2 array of individual p-n junction photodiodes, separated by a small gap of typically less than 0.05\% of the active area, as depicted in the inset of Figure~\ref{fig:set-up}. The photodiodes provide the photocurrents $I_\mathrm{A}$, $I_\mathrm{B}$, $I_\mathrm{C}$ and $I_\mathrm{D}$ which are generated when an optical beam strikes the active area. Its position-current relation can be written as:
\begin{subequations}
\begin{equation}
\label{subeq:Ix}
\frac{I_x}{I_\Sigma} = \frac{I_\mathrm{A}+I_\mathrm{D}-(I_\mathrm{B}+I_\mathrm{C})}{I_\mathrm{A}+I_\mathrm{D}+I_\mathrm{B}+I_\mathrm{C}}=\frac{\int_0^x\int_0^\infty\! |U(x,y)|^{2}\,\mathrm{d}y\mathrm{d}x}
{\int_0^\infty\int_0^\infty\! |U(x,y)|^{2}\,\mathrm{d}y\mathrm{d}x},
\end{equation}
\begin{equation}
\frac{I_y}{I_\Sigma} = \frac{I_\mathrm{A}+I_\mathrm{B}-(I_\mathrm{C}+I_\mathrm{D})}{I_\mathrm{A}+I_\mathrm{B}+I_\mathrm{C}+I_\mathrm{D}}=\frac{\int_0^y\int_0^\infty\! |U(x,y)|^{2}\,\mathrm{d}x\mathrm{d}y}
{\int_0^\infty\int_0^\infty\! |U(x,y)|^{2}\,\mathrm{d}y\mathrm{d}x},
\end{equation}
\end{subequations}
for shifts along the $x-$ and $y-$axis, respectively, with $|U(x,y)|^{2}$ the intensity of the impinging beam. To obtain the nominal beam displacement, the quadrant detector signal $I_{x,y}/I_\Sigma$ has to be normalized to the slope of this relationship curve, i.e. the calibration constant $K$. 

Typically, positional beam measurements using a quadrant detector involve a fundamental Gaussian mode profile that is cylindrically symmetric, i.e. having an isotropic profile in the cylindrical coordinate system. In that case, the quadrant detector response is most sensitive around the beam center for a small displacement $\Delta x \ll w$, and the calibration constant $K$ is derived around the beam center where the slope of positional-current relationship is linear. Previously, the quadrant detector calibration constant for LG beams as a function of $\ell$ has also been derived for small displacements around the beam center \cite{Hermosa:OptLett2011}, which is valid since LG beams have also isotropic intensity profiles. For the case of HLG beams, however, one can immediately sees from Figure~\ref{fig:HLGbeams} that the intensity profile is not cylindrically symmetric. In other words, HLG beams carrying non-integer OAM have anisotropic profiles. As will be discussed in the next paragraphs, there are two things to note when operating a quadrant detector for position measurement of anisotropic beams such as HLG beams.

First, the orientation of the HLG beam profile influences the quadrant detector response. Due to the rectangular geometry of a quadrant detector, it is most natural to align the symmetry axes of the beam with respect to the quadrant detector displacement axes, as in the case of Figure~\ref{fig:HLGbeams}. When these axes are aligned, the quadrant detector calibration constant $K$ for the displacement along the $x-$axis of HLG$_{n,m| \alpha}$ mode is also valid for the displacement along the $y-$axis of HLG$_{m,n| \alpha}$ mode.

Second, operating a quadrant detector around the HLG beam center to detect small displacement $\Delta x \ll w$ will not always give the most sensitive position measurement. This is due to the fact that for some cases, the profile cross section $|U(x,y)|^{2}$ of the HLG modes along the axis of displacement, has near zero values across one displacement axis. As an example, let us observe HLG modes for $\alpha>45^\mathrm{o}$ in Figure~\ref{fig:HLGbeams} (b). The low intensity values at the beam centre across the $y$-axis is certainly the least sensitive area to measure beam displacement along the $x-$axis. Interchangeably, the quadrant detector is least sensitive for beam displacement along the $y-$axis around the centre area of HLG modes for $\alpha<45^\mathrm{o}$. Therefore, it is important to find the region where the quadrant detector can operate with the highest sensitivity. 

\subsection{Analytical solutions of quadrant detector calibration}
Now, we derive the analytical expression for the position-current relationship of a quadrant detector for HLG beams carrying non-integer OAM when both of the symmetry axes overlap. This expression can be easily extended for an arbitrarily high order HLG$_{n,m|\alpha}$ mode. For didactic purposes, we take the example of a radial mode index $p=0$ and an azimuthal mode index $\ell=1$ (i.e. HLG$_{0,1|\alpha}$), and investigate the quadrant response for the beam displacement along the $x-$axis. By applying the distribution function of HLG$_{0,1|\alpha}$ given in Ref.~\cite{Abramochkin:JOA2004} into Equation~(\ref{subeq:Ix}), we can write the $x$-axis displacement relationship normalized to the beam radius $w$ for HLG$_{0,1|\alpha}$ to be
\begin{equation}
\frac{I_x}{I_\Sigma} = -\frac{2\sqrt{2}x}{\sqrt{\pi}w}\exp\left[-2\left(\frac{x}{w}\right)^2\right]\cos^2\left(\alpha\right)+\operatorname{erf}\left(\sqrt{2}\frac{x}{w}\right).
\end{equation}
Note that due to the symmetry axes, the same expression is found for the beam displacement of HLG$_{1,0|\alpha}$ mode along the y-axis, substituting the index $x$ with $y$. 

\begin{figure}[htbp]
\begin{center}
\includegraphics[width=12.5cm]{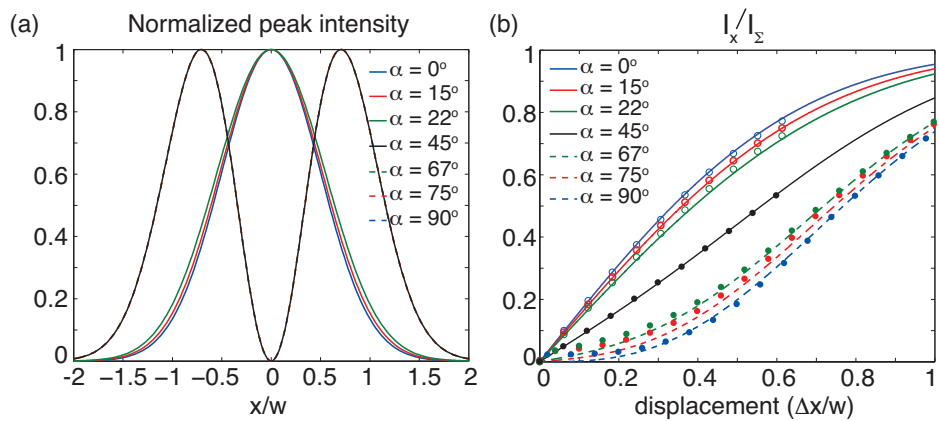}
\end{center}
\caption{\label{fig:QD-response} (a) The cross section of the HLG$_{0,1|\alpha}$ mode profile along the $x$-axis. (b) The corresponding response of a quadrant detector for beam displacement along the $x-$axis. Lines (both solid and dashed) and data points correspond to the analytical solution and experimental data, respectively.}
\end{figure}

We present the 1-D cross section profile in Figure~\ref{fig:QD-response} (a) to help visualizing the general intensity distribution of HLG$_{0,1|\alpha}$ beams. The position-current relationship curves in Figure~\ref{fig:QD-response} (b) reveal that there are different linear regions with a constant slope (calibration constant $K$) for different values of $\alpha$. The linear region shifts to a higher $x/w$ value for $\alpha>45^\mathrm{o}$, coinciding with the peak intensity cross section along the displacement axis, at around $x/w = 0.7$. For $\alpha<45^\mathrm{o}$ the beam cross-section along the $x$-axis resembles that of a Gaussian profile and the range of linearity is around the beam center. 
\begin{figure}[htbp]
\begin{center}
\includegraphics[width=12cm]{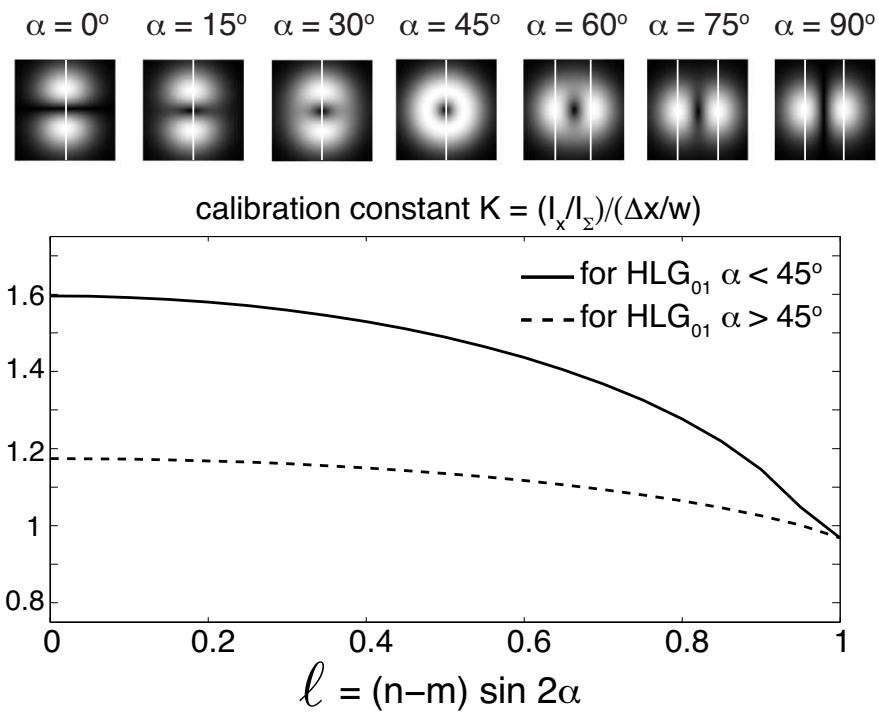}
\end{center}
\caption{\label{fig:KHLG} The calibration constant $K$ as a function of non-integer OAM $\ell$ is derived analytically for HLG$_{0,1}$ modes with varying $\alpha$. The white lines on the top row images illustrate the $x$-positions at which $K$ is derived for several $\alpha$ values.}
\end{figure}

To confirm our analytical expression, we measure the quadrant detector response for HLG$_{0,1|\alpha}$ modes, where we have used a quadrant detector from NewFocus model 2921 with an active area of $10\mathrm{mm}\times10\mathrm{mm}$. Figure~\ref{fig:QD-response} (b) shows the the match between our data (open circles and dots) and the analytical solution (solid and dashed lines). It is important to realize that there exists two values of $\alpha$ for each HLG$_{n,m| \alpha}$ mode that give the same $\ell$ values but with orthogonally oriented spatial distribution. These orthogonally oriented modes have different calibration constants $K$, as plotted in Figure~\ref{fig:KHLG}. 

To use a quadrant detector for displacement measurement of non-integer OAM beams having anisotropic profile distributions, such as HLG beams, one must pay attention to the linear range of the position-current relations, i.e. at the peak of the intensity cross section along the axis of displacement. Since the orientation of our generated HLG beam are aligned with the symmetry axes of a quadrant detector, we can easily obtain the linear range and the calibration constant $K$. This calibration procedure is particularly relevant for potential applications using HLG modes as non-integer OAM beams: in beam shifts measurements, high precision metrology, optical manipulation using tweezers or scanning near-field optical microscopy.

\section{Conclusion}
In this paper, we demonstrate a technique to generate HLG modes as non-integer OAM beams that are structurally propagation invariant and having a fixed symmetry axes for arbitrary non-integer OAM values. The experimentally demonstrated HLG beams agree with the calculation, both for the intensity profile distribution and the phase features measured with interferometric set-up.

Unlike an integer OAM beam, the phase of a HLG mode increases non-linearly along the azimuthal axis. Note that any integer OAM beam can be created from an arbitrarily higher order HLG mode having the appropriate orientation angle $\alpha$. For example, $\ell=1$ can be constructed from HLG$_{0,2|[\alpha:15^\mathrm{o}]}$, which actually produces a phase distribution that is different from that of a LG$_{0,1}$ mode. In applications such as  OAM beams shifts or optical manipulation using OAM beams, noticeable differences will occur when addressing an integer OAM value by using either LG modes or HLG modes.

For many applications using OAM beams, it is of high interest to measure accurately the beam position. Down to nanometer precision of beam displacement is typically measured using a quadrant detector. Different from previous techniques, the symmetry axes of our generated non-integer OAM beams are \textit{always} aligned to the axes of quadrant detectors; which simplifies the operation and calibration of the detector. 

We have derived the analytical expression and demonstrate experimentally the response of a quadrant detector towards the generated HLG beams. The obtained calibration constant $K$ of a quadrant detector for HLG beams agrees with Ref.~\cite{Hermosa:OptLett2011} only at integer $\ell$, where the beam profile is isotropic or cylindrically symmetric. The assumption that a quadrant detector is most sensitive at the beam center does not always hold for general astigmatic modes, i.e. HLG modes, that has an anisotropic beam profiles. 

In conclusion, we have shown that both the $\ell$ values and the HLG mode orientation play a role in the quadrant detector response. Furthermore, the anisotropic nature of HLG beams creates different regions having linear response of a quadrant detector when measuring beam positional shift. The beam positional measurement is most sensitive around the peak of the HLG mode profile. Our result can easily be extended to arbitrarily higher order HLG beams as solutions of light carrying higher order non-integer OAM.

\section*{Acknowledgement} 
This work is supported by the European Union within FET Open-FP7 ICT as part of STREP Program 255914 Phorbitech.

\end{document}